\documentclass[twocolumn,amsmath,amssymb,aps,prx,shownopacs,prx]{revtex4-1}

\usepackage{graphicx}% Include figure files
\usepackage{dcolumn}% Align table columns on decimal point
\usepackage{bm}% bold math
\usepackage{color}

\newcommand{\tp}{t'}

\newcommand{\CC}{\mathcal{C}}

\newcommand{\Tel}{T_\text{el}^*}

\begin{document}

\title{Coexistence of excited polarons and metastable delocalized states in photo-induced metals}

\author{Sharareh Sayyad} 
\author{Martin Eckstein}
\affiliation{Max Planck Research Department for Structural Dynamics, University of Hamburg-CFEL, 22761 Hamburg, Germany}

\date{\today}

\begin{abstract}
We study how polaronic states form as a function of time due to strong electron-phonon coupling, starting from a hot 
electron distribution which is representative of a photo-induced metallic state immediately after laser excitation. For this purpose 
we provide the exact solution of the single-electron Holstein model within nonequilibrium dynamical mean-field theory. In particular, 
this allows us to reveal key features of the transient metallic state in the numerically most challenging regime, the adiabatic regime, 
in which phonon frequencies are smaller than the electronic bandwidth: Initial coherent phonon oscillations are strongly damped, 
leaving the system in a mixture of excited polaron states and metastable delocalized states. We compute the time-resolved
photoemission spectrum, which allows to disentangle two contributions. The existence of long-lived delocalized states suggest 
ways to externally control transient properties of photo-doped metals. 
\end{abstract}


\maketitle

\section{Introduction}

Ultra-short laser pump-probe techniques in condensed-matter systems have opened the possibility
to generate correlated nonequilibrium phases, such as photo-induced metallic states in Mott insulators
\cite{Iwai03}, and to study their dynamics on femtosecond timescales.
On a fundamental level, seeing how correlations evolve {\em in time} can shed new light on many-body 
effects which have been investigated for decades under equilibrium conditions. A paradigm example is the 
formation of polaronic quasiparticles, i.e., the self-trapping of an electron in a lattice distortion, or ``phonon cloud''.
This phenomenon was predicted in the early days of quantum mechanics \cite{Landau1933} and has been 
thoroughly investigated for a large class of systems \cite{Devreese2009, AlexandrovBook, FehskeBook},
more recently also for ultra-cold gases and trapped ions \cite{Bruderer2007, Stojanovic2012}. 
In nonequilibrium, however, many questions related to the dynamics of systems with strong 
electron-phonon coupling remain only partially understood.

Signatures of strong electron-phonon coupling and polaronic effects in photo-excited systems have been found
for the self-trapping of excitons \cite{Tomimoto1998, Dexheimer2000, Sugita2001, Morissey2010}, in Mott 
insulators \cite{Dean2011,Novelli2014}, and organic materials \cite{Morrissey2013, kaiser2014optical, Mitrano2014}. 
A direct observation of the self-localization process was achieved by two-photon photoemission for electrons in 
surface states which couple to adsorbate layers \cite{Ge1998, Ge2000, Miller2002, Gahl2002}. 
While polaronic effects can be visible already within one phonon period after photo-excitation, it is not entirely clear 
how, and how fast, the actual polaron ground state can be reached. 
The presence of non-equilibrated polarons, on the other hand, would determine 
carrier mobilities in transient metallic states and can thus be of importance also for potential technological applications 
like ultra-fast switches. It is therefore important to pinpoint 
signatures of excited polarons, to understand  
their properties, and whether these can be controlled, e.g., by the photo-excitation process.

These questions have motivated considerable effort to understand the nonequilibrium polaron dynamics
from a theoretical perspective. A large body of work has been performed for the Holstein model \cite{Holstein1959a},
which describes tight-binding electrons coupled to an optical phonon with frequency $\omega_0$. The physical 
picture for the polaron formation process which has emerged from these studies 
suggests
two important bottlenecks: For 
large $\omega_0$, one finds long-lived beating between well-separated polaron sub-bands in the many-body 
spectrum \cite{Ku2007, Fehske2011}, while in the opposite and experimentally very relevant adiabatic regime, 
in which $\omega_0$ is smaller than the electron hopping, a semiclassical argument \cite{Emin1976, Kabanov1993} 
predicts an energy barrier between delocalized and localized states. In the present work we solve the model exactly 
in the large coordination limit to see how relaxation of high energy electrons by emission of phonons, strongly damped 
coherent oscillations, long-lived delocalized states, and trapping in excited polaron states come together in 
particular in the adiabatic limit 
and how they are
reflected in characteristic signatures of the photoemission spectrum.

Even for a single electron (the relevant limit to describe diluted polarons), the Holstein model is difficult to be solved 
in nonequilibrium, because established approaches like Quantum Monte Carlo \cite{Prokofev1998} 
cannot be used. Variants of exact diagonalization \cite{Ku2007, Fehske2011, DeFilippis2012, DeFilippis2012b, 
Vidmar2011, Matsueda2012, Golez2012} 
provide an accurate and versatile description of the electron-phonon dynamics in many regimes, but they 
rely on a cutoff of the phonon Hilbert space and become challenging in the adiabatic regime in which the 
phonon cloud involves a large number of oscillator quanta. Our work is based 
on nonequilibrium dynamical mean-field theory (DMFT) \cite{Aoki2014}, which is exact in the limit of 
large lattice coordination numbers \cite{Metzner1989}. 
In DMFT, a lattice model is mapped onto a single impurity coupled to a self-consistent 
bath. While the real-time dynamics of this impurity problem can usually be solved only approximately (see, e.g., 
Ref.~\cite{Werner2013phonon} for the Holstein model), the limit of low electron density in the Holstein model 
provides a remarkable exception. In equilibrium, the DMFT equations for this case can be written exactly in terms 
of a continued fraction for the electron Green's function \cite{Ciuchi1997}. Technically, this solution is similar to earlier 
diagrammatic approaches \cite{Sumi1974, Cini1988}, and also to the momentum averaged technique 
\cite{Goodvin2006, Berciu2006}, which have provided a solution throughout all regimes of the single-electron 
Holstein model in equilibrium (adiabatic and non-adiabatic, weak and strong coupling). These diagrammatic techniques 
rely on a  momentum-independent self-energy which is approximate in finite dimensions, but shows good agreement 
with Monte Carlo particularly in the strong-coupling regime \cite{Goodvin2011}. Here we generalize the exact DMFT 
solution of Ref.~\cite{Ciuchi1997} to the case of nonequilibrium DMFT. 

\section{Model and Methods} 

The Holstein model \cite{Holstein1959a} is defined by the Hamiltonian
\begin{align}
\label{eq:Hol1}
& H
 =
-J\sum_{\langle ij \rangle}(c_{i}^{\dagger}c_{j}+ h.c.)
+\sum_i H_\text{loc}^{(i)},
\\
\label{eq:Hol2}
&H_\text{loc}^{(i)}
=
\omega_{0} b_{i}^{\dagger}b_{i}
+g n_{i}(b_{i}+b_{i}^{\dagger}) + \epsilon_f n_i.
\end{align}
The first term in Eq.~\eqref{eq:Hol1} describes tight-binding electrons with nearest-neighbor hopping $J$ on a lattice;
 $c_{i}^{\dagger}$ and $c_{i}$ are the creation and annihilation operator of an electron on lattice site $i$,
respectively. The local part \eqref{eq:Hol2} of the Hamiltonian represents one harmonic oscillator
with frequency $\omega_0$ at each lattice site, i.e., a dispersion-less optical phonon mode,
whose coordinate $X_i=(b_i^\dagger+b_i)/\sqrt{2}$ is linearly coupled to
the electron occupancy $n_i= c_i^\dagger c_i$; $\epsilon_f$ defines the zero of the energy.
We focus on the dilute limit, where correlations between electrons are negligible,
so that expectation values of observables are proportional to the density $n_{el}=\langle c_i^\dagger c_i\rangle$ and
can obtained from the solution of the model with only one electron. The hopping $J$ is taken as a unit of energy, 
and times are measured in terms of $1/J$. The results are obtained for a Bethe lattice with a semi-elliptic 
density of states $D(\epsilon) =  \sqrt{4-\epsilon^2}/\sqrt{2\pi}$. 

To get an understanding of polaron formation in the Holstein model,
the limit of isolated lattice sites (atomic limit) is a convenient starting point \cite{MahanBook}.
In this limit, the presence of an electron on the site shifts the equilibrium position of the oscillator:
omitting site-indices for convenience, the local part of the Hamiltonian can be rewritten as 
\begin{equation}
\label{hloc-shift}
H_{\mathrm{loc}} = \frac{\omega_{0}}{2}\big[(X+X_0)^2 + P^2\big]  + (\epsilon_f-E_P)\,\hat n,
\end{equation} 
where $X=(b^\dagger+b)/\sqrt{2}$ and $P=i(b^\dagger-b)/\sqrt{2}$ are coordinate and momentum
of the oscillator, respectively, $X_0 = \sqrt{2}g/\omega_0 \hat n$, and
\begin{align}
E_P=\frac{g^2}{\omega_0},
\end{align}
is the lowering of the ground state energy which defines the bare {\em polaron binding energy}. 
In the lattice model, the energy ratio $E_P/J$ distinguishes the regimes of weak-coupling ($E_P\ll J$) 
and strong coupling ($E_P \gg J$). For strong coupling, self-localized electron
states at energy $E=-E_P$ at different sites are coupled by the hopping and form a band of delocalized polaronic states;
it's bandwidth is reduced with respect to the free bandwidth by the Frank-Condon factor, which takes into account the 
coherent motion of the lattice distortion with the electron, i.e., the overlap $|\langle 0|e^{iPX_0}|0\rangle|^2$ between 
the ground states $|0\rangle$ and $e^{iPX_0}|0\rangle$ of the oscillator and the displaced oscillator \eqref{hloc-shift} respectively. 
A second important scale for the Holstein model is the ratio $\alpha=\omega_{0}/J$, which distinguishes the adiabatic 
behavior ($\alpha \lesssim 1$), in which the phonon is slow compared to the electron, from the non-adiabatic  behavior 
($\alpha \gtrsim 1$). In the adiabatic strong-coupling regime, the number of oscillator quanta in the phonon cloud 
proliferates (in the atomic limit, $\langle b^\dagger b \rangle=g^2/\omega_0^2=E_P/\omega_0$), which makes the 
dynamics in this regime qualitatively distinct from the non-adiabatic regime. 

To study polaron formation in time, we start the simulations from an initial state in which electrons and lattice are decoupled, and the mean kinetic energy of the electron is comparable to the free bandwidth, whereas the lattice temperature 
$T_\text{latt}$ is low ($T_\text{latt} < J,\omega_0$). This initial state may be taken as a simple model for the situation 
immediately after electrons have been promoted into an empty valence band by photo-excitation from a conduction band, 
because the process of rapid inter-band excitation leaves the lattice unaffected up to a good approximation.
The precise form of the initial electron energy distribution  is not important for the subsequent dynamics
as long as it is broad on the 
scale of the bandwidth, 
and we take it to be a hot electron distribution with electron
temperature  $\Tel\sim 1-10 \,J$.

To monitor the dynamics of the model we compute the time-resolved photoemission spectrum, which can be obtained 
from the electronic Green's function. In the low density limit, the relevant propagators
for adding an electron ($\widetilde G^>$) and removing an electron ($\widetilde G^<$) are given by
\begin{align}
\label{ggtrtilde}
\widetilde G^>(t,\tp)
&=
 \frac{-i}{Z_{0}}\mathrm{Tr}_{N=0}[e^{-\beta H}c_i(t)c_i^\dagger(t')],
\\
\label{glestilde}
\widetilde  G^<(t,\tp)
&=
\frac{i}{Z_{1}n_{el}}\mathrm{Tr}_{N=1}[e^{-\beta H}c_i^\dagger(t')c_i(t)],
\end{align}
where $\text{Tr}_{N=n}$ is the trace over the $n$-electron sector, and $Z_n=\mathrm{Tr}_{N=n}[\,e^{-\beta H}]$.
(Note that we have assumed translational invariance and normalized $\widetilde G^<$ by the electron density $n_{el}$,
so that $\widetilde G^<(t,t)=i$.) The photoemission spectrum as a function of probe time $t$ and energy $\omega$ is obtained
from  $\widetilde G^<$ by partial Fourier-transform and convolution with the envelope $S(t)$ of the probe pulse
\cite{FreericksKrishnamurthyPruschke2009},
\begin{equation}
\label{trpes}
I(\omega,t)=
\int
\frac{dt_1dt_2}{2\pi i}
 \,S(t_1)S(t_2)\,e^{i\omega(t_1-t_2)} \widetilde G^{<}(t+t_1,t+t_2).
\end{equation}
In equilibrium, $\widetilde G^{<}(t,t')$ is translationally invariant in time, so that $I(\omega)$ is given by the convolution 
\begin{align}
\label{eqpes}
I(\omega) = \int d\omega'  A^{<}(\omega-\omega') |\tilde S(\omega')|^2,
\end{align}
of the power spectrum 
$|\tilde S(\omega)|^2=|\int \!dt\, e^{i\omega t }S(t)|^2/2\pi$ 
of the probe pulse with the occupied density of states, $A^{<}(\omega) = 
(1/2\pi i)\int  dt \, e^{i\omega t} \widetilde G^{<}(t,0)$. In addition to the photoemission spectrum, we will compute time-local 
observables, i.e., the kinetic energy per site,  $E_\text{kin}(t) = -J\sum_{\langle ij\rangle } \langle c_i^\dagger c_j \rangle/Ln_{el}$,
as well as the average number of oscillation quanta in the phonon cloud (i.e., at a site occupied by an electron),
$N_{ph}(t)= \langle n_i b_i^{\dagger}b_i\rangle/n_{el}$ (the expectation values are translationally invariant
and normalized by the electron density).

We compute the dynamics of the Holstein model using the nonequilibrium generalization of DMFT \cite{Aoki2014}.
In the limit of low density, the solution can be made exact, yielding both Green's functions \eqref{ggtrtilde} and 
\eqref{glestilde}. In equilibrium \cite{Ciuchi1997}, computing the propagator $\widetilde G^>$ is sufficient, because 
$\widetilde G^>$ and $\widetilde G^<$ are related by a fluctuation dissipation relation. For the nonequilibrium case, 
we thus have to reformulate the equations of Ref.~\cite{Ciuchi1997} in real-time and provide additional equations 
for $\widetilde G^<$ (or equivalently, one set of equations on the Keldysh contour). The resulting equations are 
Volterra integral equations whose numerical solution is controlled by the maximum number $N_\text{max}$ of 
phonons on each site; 
the computational effort increases however only linearly with $N_\text{max}$, 
so that we can obtain converged 
results 
with $N_\text{max}=50$ for several tens of hopping times.
To keep the presentation concise, the detailed 
formalism is explained in the appendix \ref{ap1}.

\section{Results} 

\begin{figure}[tbp]
%\centerline{ \includegraphics[width=0.99\columnwidth]{Fig1_1.eps}}
%\centerline{ \includegraphics[width=0.99\columnwidth]{Fig1_2.eps}}
\centerline{ \includegraphics[width=0.99\columnwidth]{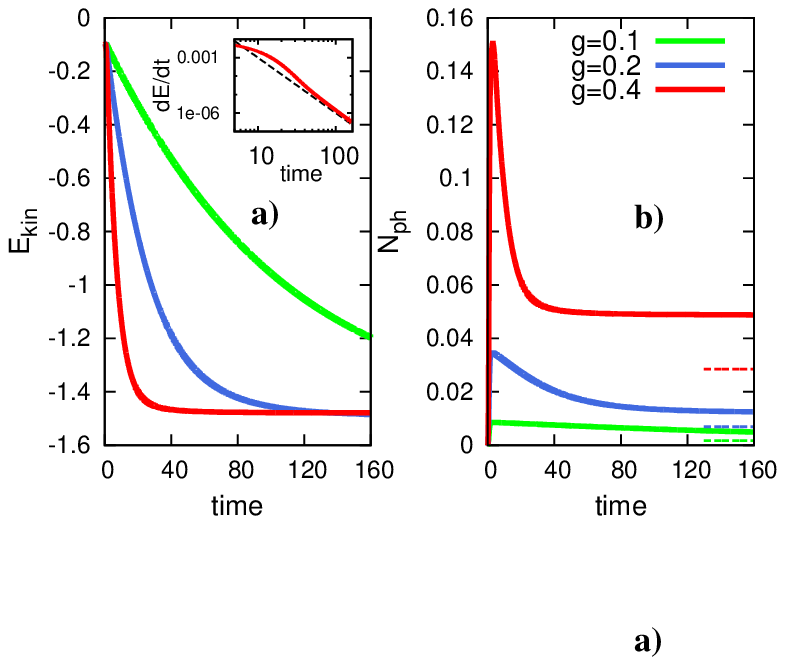}}
\centerline{ \includegraphics[width=0.99\columnwidth]{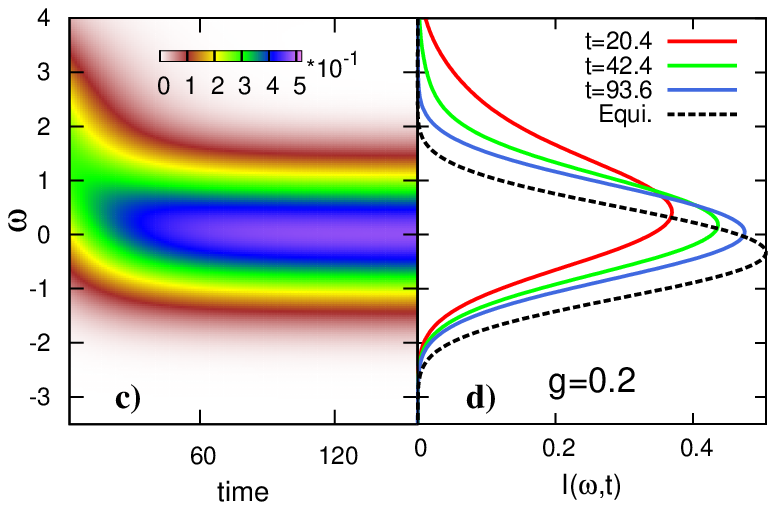}}
 \caption{
Relaxation in the weak-coupling regime.  {\bf a)} Time-evolution of the kinetic energy for
three values of the coupling ($T_{\mathrm{latt}}=0.1$, $\Tel=10$, $\omega_{0}=1$). The inset
shows the power law behavior of $dE_{kin}/dt$ for $g=0.4$; the red line are data, the 
dashed black line is a power law $\sim1/t^3$.  {\bf b)} Time-evolution of the 
average phonon number for the same parameters. The horizontal dashed lines indicates
the corresponding values of $N_{ph}$ in thermal equilibrium at $T=T_\text{latt}$.  {\bf c)} 
and {\bf d)} Time-resolved 
photoemission spectrum $I(\omega,t)$
for $g=0.2$. The spectrum is obtained from Eq.~\eqref{trpes},
using a Gaussian probe pulse $S(t)\propto\exp(-t^{2}/2\delta^{2})$ with duration $\delta=3$.}
\label{fig:Ekin_weak}
\end{figure}

\subsection{Weak coupling regime}

The weak-coupling regime is rather well described by rate equations (see below),
which can capture the cooling of the initial hot electron state by emission of phonons. 
Nevertheless it is illustrative to
look at the corresponding DMFT solution, to contrast the 
behavior
for strong-coupling below. Figure~\ref{fig:Ekin_weak}a 
and b show the relaxation of the kinetic energy and the phonon number $N_{ph}$ for various coupling strength $g$.
After a short transient, the time-evolution of both quantities follows a monotonous relaxation, which becomes faster 
with increasing coupling strength. Similarly, the relaxation can be seen in the time-resolved photoemission spectrum 
(Fig.~\ref{fig:Ekin_weak}c). At early times, the occupied density of states reflects the initial hot electron state and is smeared
over the full band. 
(In the uncorrelated equilibrium state, the occupied density of states is 
$A^<(\omega) \propto D(\omega)e^{-\omega/{\Tel}}$.) 
Subsequently, electrons reduce their kinetic energy by the emission of phonons, and 
spectral weight is concentrated closer to the lower band edge. 

\begin{figure*}[tbp]
\centering
\includegraphics[width=\columnwidth]{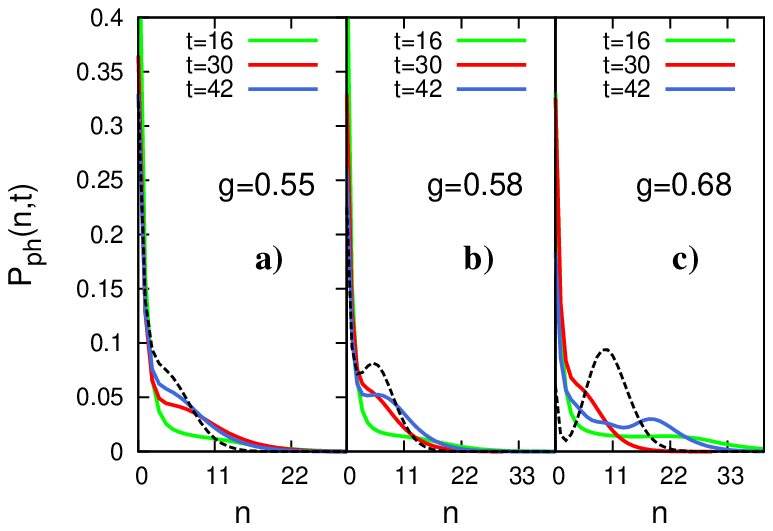}
\includegraphics[width=\columnwidth]{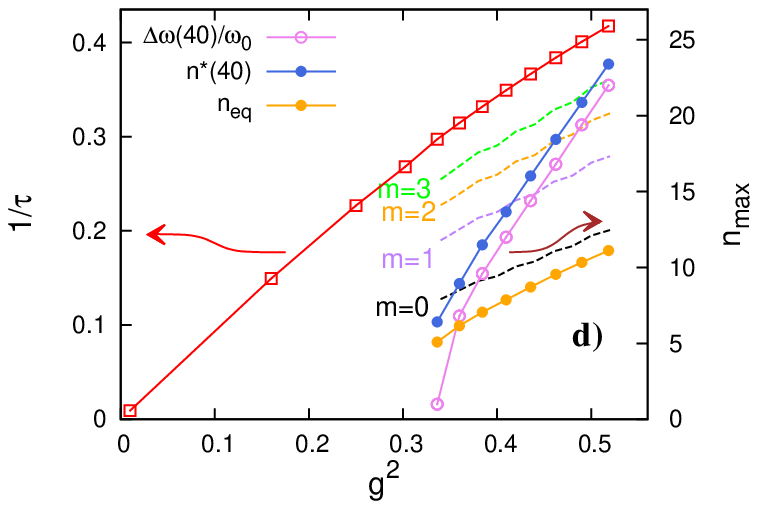}
\caption{Phonon number distribution and polaron crossover. {\bf a)} to {\bf c)} 
$P_\text{ph}(n)$  at $\omega_{0}=0.2$  for different couplings and times ($T_{\mathrm{latt}}=0.1$, $\Tel=10$). 
The dashed black line corresponds to the equilibrium state at temperature $T_\text{latt}$. 
{\bf d)} The position of the maxima in $P_\text{ph}(n)$ for equilibrium ($n_{eq}$, orange filled circles) and 
at time $t=40$ 
(blue filled circles, see right vertical axis). 
Open symbols show the ratio $\Delta \omega/\omega_0$ 
at the same time, where $\Delta \omega$ is the splitting of the two peaks in the photoemission spectrum. 
Dashed lines labelled
 $m=0,1,2,3$ show the position of the maximum of the distribution functions of the 
displaced oscillator in it's $m$th eigenstate [c.f.~Eq.~\eqref{laguerre}) with $\gamma=g/\omega_0$, the 
maximum with the largest $n$ is shown]. The red curve with square symbols
(left vertical axis)
shows the relaxation time $1/\tau$ of the high-energy part of the photoemission spectrum
(see main text).}
\label{fig:dist_strong1}
\end{figure*}

For weak electron-phonon coupling, relaxation phenomena at long times are captured by
a kinetic equation \cite{Mahan1987}, which is also in agreement with exact diagonalization studies
\cite{Ku2007, Golez2012}. For low lattice temperature ($T_\text{latt}\ll \omega_0$), an electron with 
band energy $\epsilon$ 
can only emit phonons, at a rate 
determined by the coupling $g$ and the density of (final) 
states,
\begin{equation}
\label{trelax}
\frac{1}{\tau(\epsilon)}  = g^2 D(\epsilon-\omega_0).
\end{equation}
This result is obtained from Fermi's golden rule, or equivalently,
the imaginary part of the equilibrium self-energy $\text{Im} \Sigma(\epsilon+i0)$.
The $g^2$-dependence of the relaxation time is indeed confirmed by the DMFT results when one fits
the time-dependence of the photoemission spectrum $I(\omega,t)$ in a certain energy window with a simple
exponential function $A \exp(-t/\tau) +C$ (this will be analyzed further below, see the curve $1/\tau$ in 
Fig.~\ref{fig:dist_strong1}d). Furthermore, from Eq.~\eqref{trelax} one sees that a thermal equilibrium state 
can never be reached, because the density of states vanishes if the  final energy $\epsilon-\omega_0$ is below 
the lower band edge. This phase-space effect can be seen explicitly in our data: At long times,
the time-resolved photoemission spectrum remains shifted with respect to the spectrum of the equilibrium state
at temperature $T=T_\text{latt}$ (see dotted horizontal lines in Figs.~\ref{fig:Ekin_weak}d).
 
Finally, we note that due to the energy-dependent relaxation time, the long-time asymptotic behavior of averaged 
quantities is not necessarily exponential. This can be seen for the kinetic energy: For a density of states 
$D(\epsilon) \propto \sqrt{\epsilon-E_0}$  with a van-Hove singularity at the lower band edge $E_0$ (as for a 
three-dimensional lattice, or the semi-elliptic density of states used here), the rate Eq.~\eqref{trelax} 
implies a power-law long-time asymptotic behavior of $E_\text{kin}$ 
with $d E_\text{kin}/dt  \sim t^{-3}$. (For a 
one-dimensional density of states, one would expect an exponential decay \cite{Golez2012}.) This behavior 
is 
observed 
in the numerical data (see Fig.~\ref{fig:Ekin_weak}a, inset), which is a nice confirmation of the 
rate equation analysis.

\begin{figure}[tbp]
\centering
\includegraphics[width=\columnwidth]{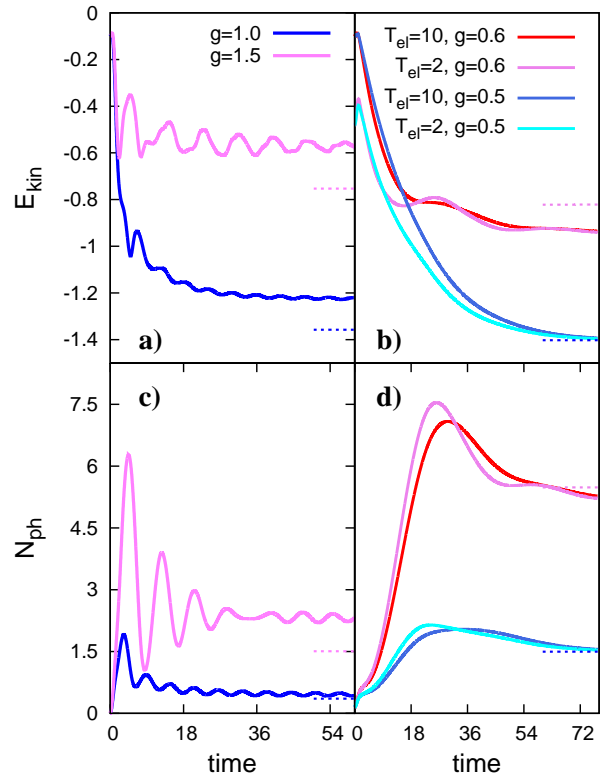}
\caption{Relaxation of $E_{kin}$ and $N_{ph}$ at strong and intermediate coupling. {\bf a)} and {\bf c)}
Non-adiabatic regime ($\omega_0=1$), for $E_P=1$ ($g=1$) and $E_P=2.25$ ($g=1.5$), 
and $T_\text{latt}=0.2$ and $\Tel=10$. {\bf b)} and {\bf d)} Adiabatic regime ($\omega_0=0.2$), for 
$E_P=1.25$ ($g=0.5$) and $E_P=1.8$ ($g=0.6$),  and initial conditions $\Tel=1$ and $\Tel=2$. 
Horizontal dashed lines indicate expectation values of the respective quantities in equilibrium at $T=T_\text{latt}$.
}
\label{fig:Ekin_strong1}
\end{figure}

\subsection{Strong coupling regime: Overview}

\begin{figure}[tbp]
\centering
\includegraphics[width=\columnwidth]{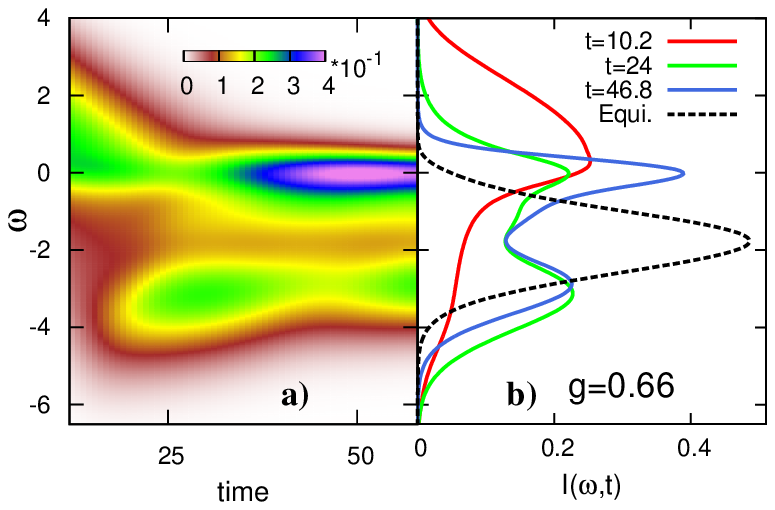}
\includegraphics[width=\columnwidth]{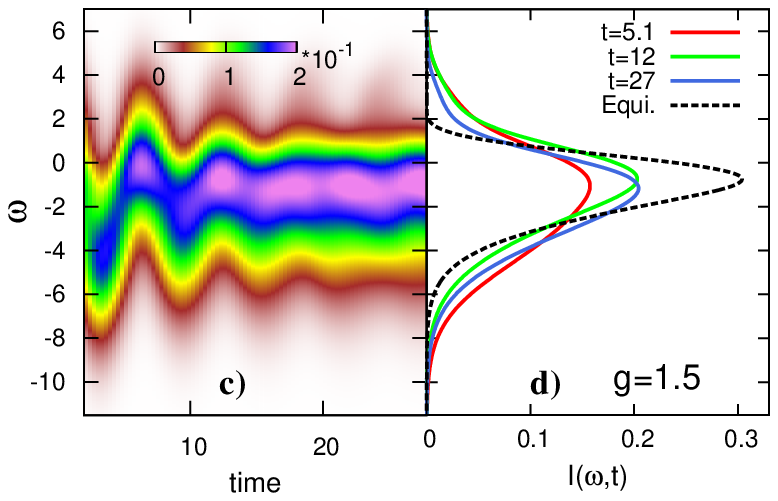}
\caption{Time-resolved photoemission spectrum $I(\omega,t)$ at strong coupling.
{\bf a)} and {\bf b)} Adiabatic regime: $\omega_0=0.2$, $g=0.66$ ($E_P=2.18$), $T_\text{latt}=0.1$, $\Tel=10$.
The spectrum is computed from Eq.~\eqref{trpes} with a Gaussian probe pulse $S(t)\propto\exp(-t^2/2\delta ^2)$
and a probe pulse duration $\delta=3$, smaller than the oscillation period $2\pi/\omega_0$. The right panel 
{\bf b)} shows the spectrum at selected times, and a comparison to the equilibrium spectrum at $T=T_\text{latt}$ 
(black dashed line); the energy zero $\epsilon_f$ is fixed such that $\omega=0$ is the lower edge of the 
free band. {\bf c)} and {\bf d)} Similar to upper panels, for a comparable value of the polaron binding $E_P$ in
the non-adiabatic regime: $\omega_0=1$, $g=1.5$ ($E_P=2.25$), $T_\text{latt}=0.1$ $\Tel=10$. Probe pulse duration $\delta=1$.}
\label{fig:PES_strong1}
\end{figure}

In the remainder of this paper we focus on the intermediate and strong coupling regime, where small polarons 
are formed in equilibrium. Figure~\ref{fig:Ekin_strong1} shows the relaxation of $E_{kin}$ and $N_{ph}$ for 
couplings $E_P\approx1$ to $E_P\approx2$, and phonon frequencies $\omega_0=0.2$ and $\omega_0=1$
in the adiabatic and non-adiabatic regime, respectively. The sudden coupling of the electron
and phonons leads to coherent oscillations, which are more pronounced
for large $\omega_0$.  Furthermore, the absolute value of the kinetic energy becomes smaller with increasing $g$,
indicating a stronger localization of the carriers, and $N_{ph}$ shows a pronounced enhancement of the phonon cloud.
These effects provide a first glance at the crossover from intermediate to strong coupling. 
A further analysis of the photoemission spectrum (Fig.~\ref{fig:PES_strong1}) will show that 
the observed dynamical results from a mixture of two different relaxation path, involving either 
delocalized and localized states.

In the adiabatic case, $\omega_0=0.2$ (Figs.~\ref{fig:PES_strong1}a and b), we can distinguish several characteristic
features in the photo\-emission spectrum: (i) A rapid decay of the weight at high energies ($\omega \gtrsim 1$, $t\lesssim 20$), 
starting from the broad distribution of the initial hot electron state. (ii) Buildup of spectral weight far below the lower edge 
of the free band (around $\omega=-3$) within less than one period $2\pi/\omega_0$,  and a beating of weight between 
this region and $\omega\approx 0$ at the frequency $\omega_0$. Finally, (iii), even though the oscillations are 
damped, the spectrum is still different from the spectrum in the thermal state at temperature $T=T_\text{latt}$ (dashed line in panel b), 
and displays two peaks instead of a single polaron band. Other than at weak-coupling, the differences between transient 
and equilibrium spectra occur on energy scales considerably larger than $\omega_0$. 
Spectra for the non-adiabatic regime ($\omega_0=1$) are shown in Figs.~\ref{fig:PES_strong1}c and d:
Coherent oscillations are reflected in a rigid-like shift of the occupied density of states, and a two-peak 
structure of the transient state is not observed. 

To develop a physical understanding of these observations, we will perform an analysis in two directions: a comparison 
to the spectrum of an isolated site will allow us to single out characteristic spectral signatures of (excited) polaron states 
and show how they reflect the structure of the phonon cloud, and a momentum-resolved spectrum will distinguish 
contributions from polarons and delocalized electrons. 

\begin{figure}[tbp]
\centering
\includegraphics[width=\columnwidth]{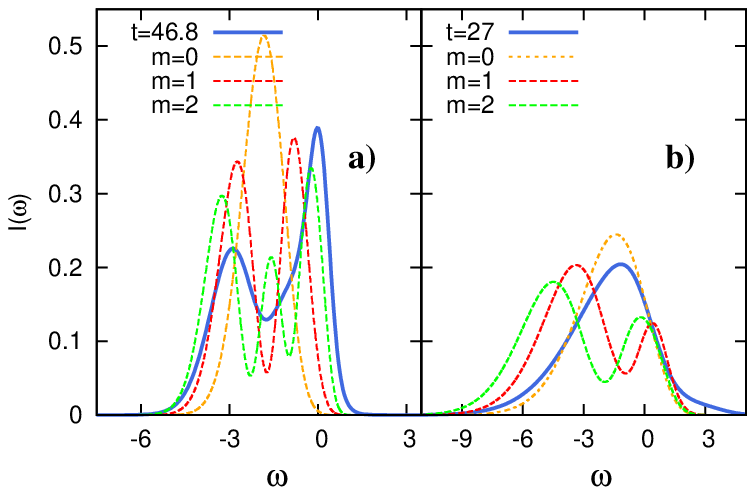}
\includegraphics[width=\columnwidth]{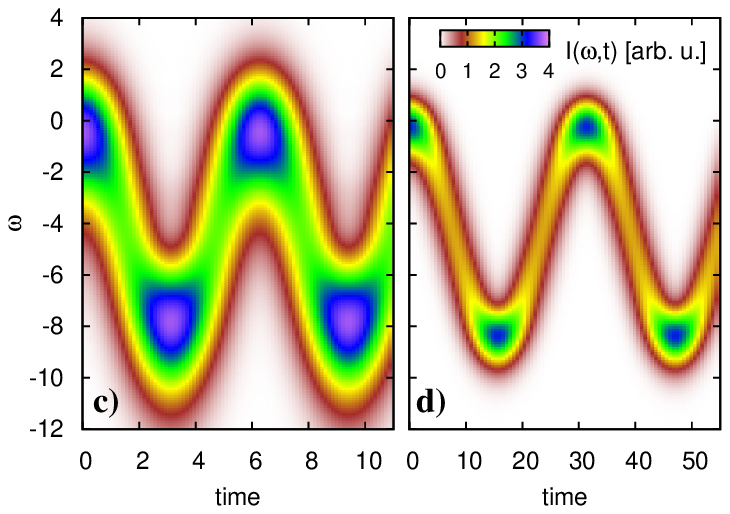}
\caption{Photoemission spectrum for the atomic limit. {\bf a)} and {\bf b)} 
Time-{\em in}dependent spectra,
assuming the initial polaron is in the ground state ($m=0$, see Eq.~\eqref{alesequiatom}) of the 
displaced
oscillator \eqref{hloc-shift},
or in an excited state $m=1,2$ [c.f.~Eq.~\eqref{laguerre}]. Parameters are like in Fig.~\ref{fig:PES_strong1}: $\omega_0=0.2$, $g=0.66$,
probe pulse duration $\delta=3$ for panel {\bf a)} and $\omega_0=1$, $g=1.5$, probe pulse duration $\delta=1$
for panel {\bf b)}. Blue solid line is the spectrum taken from Fig.~\ref{fig:PES_strong1}. 
Note that the energy zero $\epsilon_f$ for the spectra in the atomic limit is adapted to account 
for the difference between the polaron binding energy in the lattice and a the isolated site.
{\bf c)}
and {\bf d)} Photoemission spectrum after a sudden switch-on of the coupling $g$ [obtained from Eqs.~\eqref{trpes}
and \eqref{gquench}], for the same parameters as {\bf a)} and {\bf b)}, respectively.}
\label{fig:alimit}
\end{figure}

\subsection{Atomic limit and spectroscopic signatures of excited polarons}

In the atomic limit, the Holstein model can 
be solved analytically, both in and out of equilibrium, using a Lang-Firsov 
transformation \cite{MahanBook} or it's time-dependent generalization \cite{Werner2013phonon}.  Details of the solution are 
summarized in Appendix \ref{ap2}. In the ground state, the 
polaron
corresponds to the displaced oscillator 
[Eq.~\eqref{hloc-shift} with $n=1$], and the occupied density of states is given by a set of delta-peaks,
\begin{align}
\label{alesequiatom}
A^<(\omega) = \sum_{n=0}^\infty P(n) \,\delta(\omega - E_P - n\omega_0),
\end{align}
where the weights $P(m)$ are given by the phonon number distribution in the polaron state. This result has 
an intuitive understanding: photoemission removes an electron from the bound state at energy $-E_P$
and transfers the oscillator into it's excited state $|n\rangle$ with a probability which is 
given by the overlap  of $|n\rangle$ and the oscillator state $|\psi\rangle$ {\em before} removing the electron,
$|\langle n |\psi\rangle|^2=P(n)$.
At zero temperature, $|\psi\rangle = e^{iX_0 P} | 0\rangle$ is the ground state of the oscillator \eqref{hloc-shift} 
with $X_0=\sqrt{2}g/\omega$, and $P(n)=e^{-\gamma^2}\gamma^{2n}/n!$ is a Poisson distribution with 
mean $\gamma^2 = g^2/\omega_0^2$. The corresponding photoemission spectrum, Eq.~\eqref{eqpes}, 
already matches the  lattice result quite accurately for the parameters of Fig.~\ref{fig:PES_strong1}, 
as shown by the curves 
labelled $m=0$ 
in Figs.~\ref{fig:alimit}a and b.
It is thus worthwhile to take the isolated site also as a starting point to analyze the peculiar double 
peak spectra of the non-thermal state after dephasing of oscillations transient state at $\omega_0=0.2$. 
(The dephasing of oscillations is studies in more detail in Sec.~\ref{sec-oscillations} below.)

At first sight, 
one may assume that
a peak in $I(\omega,t)$ which is shifted several multiples of $\omega_0$ 
with respect to the ground state polaron implies a highly excited state. We will now argue, however, that 
the two-peak structure of the spectrum in the adiabatic case can be taken as the characteristic signature 
of a low lying excited polaron state. For this purpose we compute the photoemission spectrum for an 
isolated site, assuming that the displaced oscillator is initially in it's $m$th excited eigenstate. In this 
case Eq.~\eqref{alesequiatom} still holds, with 
the phonon excitation energy $n\omega_0$  
in the delta function replaced by $(n-m)\omega_0$. The phonon distribution function of the exited state,
$P_{m}(n) \equiv | \langle m |e^{iPX_0}| n\rangle |^2 $, is given by 
\begin{align}
\label{laguerre}
P_{m}(n+m) 
= 
P_0(n)  \frac{n! m!}{(n+m)!} L_{m}^{(n)}(\gamma^2)^2,
\end{align}
where $P_0(n) = e^{-\gamma^2} \gamma^{2n}/n!$ is the Poisson distribution of the ground state 
($\gamma=g/\omega_0$),  and $L_{m}^{(n)}(x)$ is a generalized Laguerre polynomial (see Appendix \ref{ap2}).
In particular, we have $L_{1}^{(n)}(x) = n+1 - x$, i.e., the distribution function $P_1(n)$ is suppressed 
at $n=\gamma^2$ (close to the maximum $\gamma^2$ of $P_0$), which implies a double peak. 
In general the $m$th polynomial has $m$ zeros, reflecting the probability distribution 
function of the oscillator coordinate. A comparison of these excited state spectra with the time-dependent 
spectra of the lattice model 
shows that the splitting of the two peaks in $I(\omega,t)$  (Fig.~\ref{fig:alimit}a) or the width of the distribution
(Fig.~\ref{fig:alimit}b) after the decay of the oscillations
is well in agreement with the fact that a low lying excited polaron state ($m=0,1,2$) is reached.
The main difference to the lattice result is a strong enhancement of the peak around $\omega=0$ 
in the adiabatic case, which will be analyzed in Sec.~\ref{sec:delocaloized} below.

Because in the atomic limit the photoemission spectrum reflects the number distribution function in the phonon cloud,
it is interesting to analyze $P(n)$ directly in the lattice model and see whether a similar relation can be established.
The phonon-number distribution in the lattice, which is defined by the translation-invariant correlation function
\begin{equation}
 \label{dist_ph}
P_{\mathrm{ph}}(n,t)=\frac{1}{Ln_{el}}\sum\limits_{i}\langle n_{i}\delta_{b_{i}^{\dagger}b_{i},n}(t)\rangle,
\end{equation}
is plotted in Fig.~\ref{fig:dist_strong1} for various coupling strength in the adiabatic limit. Initially (at time zero, not shown), the 
distribution is a Boltzmann distribution $P_{\mathrm{ph}}(n,0)\propto e^{-n\omega_0/T_\text{latt}}$. In the equilibrium state at coupling $g$ 
(dashed lines), the formation of a polaron is indicated by a peak at finite $n=n_{eq}$, which approaches the Poisson result 
$n_{eq}=g^2/\omega_{0}^2=E_P/\omega_0$ for large $g$, see Fig.~\ref{fig:dist_strong1}d. 
The real-time data (solid lines in Fig.~\ref{fig:dist_strong1}a-c) show an initial increase of phonon numbers (phonon states up to $n=50$ 
must be kept to simulate the dynamics in this regime). For the weaker coupling case (Fig.~\ref{fig:dist_strong1}a), $P_{\mathrm{ph}}(n,t)$ 
then evolves towards the equilibrium distribution. For couplings beyond a crossover scale $g\approx0.58$ ($g^2=0.336$, $E_P=1.68$), 
where the polaron peak forms in equilibrium, a maximum 
$n^*$ which is shifted with respect to $n_{eq}$ appears in addition to the zero-centered distribution
(Figs.~\ref{fig:dist_strong1}b,c). 
Comparison of $n^*$ with the position of the maximum of the distribution of the excited polaron 
states [Eq.~\eqref{laguerre}] for $m=0,2,3$ also confirms the previous finding that the polaron 
is transferred into an a low-lying excited state. 
A similar 
characterization of excited polaron states 
by their number distribution has also been discussed for an isolated Holstein impurity \cite{Fehske2011}.

The relation \eqref{alesequiatom} in the atomic limit would imply that the separation of the two maxima 
$n=n^*$ and $n=0$ in $P_\text{ph}$ is related to the separation $\Delta \omega$ of two peaks in the 
photoemission spectrum $I(\omega,t)$ by $\Delta \omega /\omega_0 = n^*$ (up to the energy resolution 
of the probe pulse). This relation 
indeed holds quite accurately in the lattice, see Fig.~\ref{fig:dist_strong1}d: 
the position of the maxima  $n^*$ at large time ($t=40$) depends on coupling and time, 
but it quite accurately matches the value $\Delta \omega (t)/\omega_0$ (open and filled 
blue  circles in symbols in Fig.~\ref{fig:dist_strong1}d). 
Hence the photoemission spectrum is a good measure for the phonon cloud also in the lattice model. 
In particular we note that in the adiabatic case excited polarons appear generically for couplings 
beyond crossover scale for polaron formation in equilibrium, and since the splitting $\Delta \omega$ is 
of the order of $E_P$ rather than the small scale $\omega_0$, this feature could be taken to monitor 
the polaron crossover in experiment. 
On the other hand, it is interesting to see that no signature of the crossover is seen in the behavior of 
high-energy electrons. For this we integrate the spectrum $I(\omega,t)$ over the high-energy part 
($2\le \omega \le 6$ in Fig.~\ref{fig:PES_strong1}a) and fit the result with an exponential function 
$A \exp(-t/\tau) +C$. The relaxation rate $1/\tau$ is a smooth function and almost linear with of 
$g^2$ over the whole crossover regime (see red line in Fig.~\ref{fig:dist_strong1}d).

\begin{figure}[tbp]
\centering
\includegraphics[width=0.99\columnwidth]{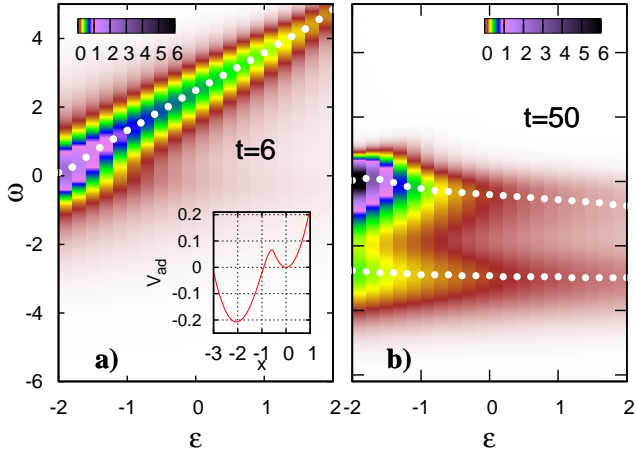}
\caption{Momentum-resolved photoemission spectrum $I(\bm k,\omega,t)$ for two different times
as a function of the electron dispersion $\epsilon_{\bm k}$, in the adiabatic case (same parameters as 
Fig.~\ref{fig:PES_strong1}a and b).
Dotted lines show the location of the maximum intensity as a function of $\omega$. 
The inset in {\bf a)} shows the adiabatic potential for $g=0.66$ and $\omega_0=0.2$
(see text).
}
\label{fig:erpes}
\end{figure}

\subsection{Disentangling free and bound states}
\label{sec:delocaloized}

We now focus on the marked asymmetry of the two peaks in the transient spectra, Fig.~\ref{fig:PES_strong1}b.
Because the peak at higher energy also roughly coincides with the energy of the lower band edge in the free band,  
one may assume that the additional weight of the peak at higher energy is due to a contribution from delocalized 
states. 
To confirm this picture, we look at the momentum-resolved photoemission spectrum 
$I(\bm k,\omega,t)$, to show that the asymmetric contribution is localized in $\bm k$. 
$I(\bm k,\omega,t)$ is obtained from Eq.~\eqref{trpes} by replacing the local Green's function  with the momentum-resolved Green's 
function 
$\widetilde G_{\bm k}^<(t,t') = i \text{Tr}_{N=1} [e^{-\beta H}c_{\bm k}^\dagger(t') c_{\bm k}(t) ]/Z$. 
 With a momentum-independent 
self-energy, dependence on $\bm k$ appears only via the electron dispersion $\epsilon_{\bm k}$, which extends 
from $-2$ to $2$ for the semi-elliptic density of states. 
The local spectrum is simply $I(\omega,t)=\int d\epsilon \,D(\epsilon) I(\epsilon,\omega,t)$. 

In Fig.~\ref{fig:erpes}, $I(\epsilon_{\bm k},\omega,t)$ is plotted for two different times. 
At early time one observes one maximum $\omega_1(\epsilon_{\bm k})$ in $I(\epsilon_{\bm k},\omega,t)$   
for each $\epsilon_{\bm k}$ (see white dotted line in Fig.~\ref{fig:erpes}a). The linear relation 
$\omega_1 \sim \epsilon_{\bm k}$ still reflects the behavior of free electrons. At later times, a 
flat band with two maxima $\omega_1(\epsilon_{\bm k})$ and $\omega_2(\epsilon_{\bm k})$
appears which reflects the polaron states (white dotted lines in Fig.~\ref{fig:erpes}b).  The ratio of the 
two maxima, $I(\epsilon,\omega_1(\epsilon),t)/I(\epsilon,\omega_2(\epsilon),t)$, is
however strongly enhanced at $\epsilon=-2$; it is $25.06$, $0.92$, and $0.597$ for 
$\epsilon=-2$, $0$, $2$, respectively. This confirms that 
the asymmetry of the two peaks in the ${\bm k}$-integrated spectrum $I(\omega,t)$ indeed 
comes mainly from the region $\epsilon_{\bm k}=-2$, and thus may be assigned to an
additional contribution from delocalized states, which could not be disentangled from 
the upper polaron peak by the energy-resolved spectrum alone.

The presence of metastable delocalized states has long been predicted from semiclassical arguments
\cite{Emin1976, Kabanov1993} from the existence of a potential energy barrier between delocalized
and polaron states in the adiabatic potential $V_{ad}(x)$. In high-dimensions \cite{Ciuchi1997}, the 
latter is given by the sum of the classical energy cost $\omega_0 x^2/2$ for displacing the oscillator at 
one lattice site, and the corresponding lowering of the ground state due to the impurity with 
potential $\sqrt{2}gx$. Since the electronic ground state energy 
is not
lowered if the impurity 
potential lies within the bandwidth, there is always an energy cost for creating small distortions, 
and thus an energy barrier for bringing the system into a self-trapped state. In infinite-dimensions,
$V_{ad}(x)$ can be computed analytically \cite{Ciuchi1997}.  In weak coupling, $V_{ad}$ slightly 
deviates from the zero-centered harmonic oscillator. 
A second minimum in $V_{ad}$ appears for $E_{P}>1.28\equiv E_{P}^{(1)}$, and
becomes the global minimum for $E_{P}>1.68\equiv E_{P}^{(2)}$, see inset Fig.~\ref{fig:erpes}b. Note 
that the scale $E_{P}^{(2)}$ is nicely in agreement with the crossover scale $g=0.58$ in Fig.~\ref{fig:dist_strong1}, 
beyond which we observe the formation of excited polarons. The 
global
minimum describes the ground 
state properties of the localized state, and the local minimum at $x=0$ corresponds to a delocalized state 
in the semiclassical picture.

\subsection{Coherent oscillations}
\label{sec-oscillations}

In this section we will finally discuss the initial coherent oscillations which follow the coupling 
of the electrons to the lattice and the resulting sudden displacement of the oscillator zero.
In the non-adiabatic regime, oscillations are reflected in a rigid-like shift of the band 
(Fig.~\ref{fig:PES_strong1}c). One can see that this is the behavior expected for a 
single oscillator:
In the atomic limit, the Green's function for a sudden switch-on of the coupling 
can be obtained exactly; it is
related to the time-translationally invariant equilibrium one
[$\widetilde G^<_{eq}(t)=i\int d\omega\, e^{-i\omega t} A^<(\omega)$, with Eq.~\eqref{alesequiatom}] 
by a simple time-dependent factor (see Appendix \ref{ap2}),
\begin{align}
\label{gquench}
\widetilde G^<(t,t') 
&=
\widetilde  G^<_{eq}(t-t') Q(t)Q^*(t'),\\
Q(t) &= \exp[2i g^2/\omega_0^2 \sin(\omega_0t)].
\end{align}
In the photoemission spectrum, Eq.~\eqref{trpes}, the oscillating factor $Q(t)$ roughly acts like a shift of 
the probing frequency $\omega$ by $2E_P \cos(\omega_0t)$ when the probe pulse is shorter than 
$2\pi/\omega_0$, so that the $\sin(\omega_0(t+t_1))\approx\sin(\omega_0t)+t_1\omega_0\cos(\omega_0t)$ 
in $Q(t)$. The resulting photo\-emission spectrum is shown in Figs.~\ref{fig:alimit}c and d. 
(Longer pulses, which average over many cycles, would lead to time-independent bands split by $\omega_0$.)

From the comparison of Fig.~\ref{fig:PES_strong1} with Figs.~\ref{fig:alimit}c and d it is apparent
that only in the non-adiabatic regime does the lattice result reflect the coherent oscillations 
found in the atomic limit. This shows a qualitative difference between the two 
regimes. In the adiabatic regime, the same bare polaron binding $E_P$ corresponds to a 
larger number of phonon energy quanta. An electron can thus easily emit several phonons 
to neighboring sites, so that vibrational dephasing occurs already on the timescale of one 
phonon-period. In the non-adiabatic regime, in contrast, the total excitation energy corresponds 
to very few oscillator quanta right from the beginning, so that emission of phonons is restricted 
by phase space effects and the system remains in long-lived beating oscillations, which is in 
agreement with results from 
exact diagonalization 
\cite{Ku2007,Fehske2011}.

\section{Conclusion}

In conclusion, we have obtained the numerically exact solution of the single-electron 
Holstein model within nonequilibrium DMFT.  The results provide a comprehensive 
picture how an excited ``hot'' electron distribution relaxes due to optical phonons, 
both at  weak and strong coupling, and in the adiabatic and non-adiabatic 
regimes. Most important are the results for small phonon frequencies (adiabatic regime) and 
strong coupling, where polaronic states are expected in equilibrium. After a quick dephasing 
of initial coherent oscillations, the system reaches a state in which excited polarons coexist 
with metastable delocalized states. While we cannot resolve the final relaxation to the ground 
state (the time range of our simulations extends to several phonon periods), the observed 
transient features are expected to be important for a photo-induced metallic state at 
strong-electron-phonon coupling. (In fact, in real systems the lifetime of the entire photo-induced 
state may be shorter than the final equilibration time.) 

Moreover, we discuss how the  photoemission spectrum reflects properties of the phonon cloud 
and can thus be used to characterize the transient state: Excited polarons lead to a characteristic 
double-peak structure of the almost flat  (i.e., weakly momentum dependent) polaron band. 
Delocalized states, on the other hand, can be identified because their distribution is peaked in 
momentum space. Nonequilibrium polarons and metastable delocalized states appear beyond 
a well-defined polaron crossover scale. At the same time, no signature of the crossover is seen 
in the relaxation behavior of high-energy electrons. This suggest that the high-energy 
relaxation rates can be used in experiment to estimate the coupling by a 
analysis in terms of Fermi Golden rule \cite{Sentef2013} even in the regime 
where small polarons are formed.

As far as a comparison is possible, our results are in qualitative agreement with earlier predictions, 
and with results for low-dimensional systems: A  beating between excited polaron states in 
the non-adiabatic case is in agreement with exact diagonalization results for one dimension 
\cite{Ku2007, Fehske2011}. The dynamics of the strong coupling adiabatic regime most difficult 
to describe in a quantum mechanical lattice calculation. A barrier for relaxation from delocalized 
states to self-trapped states was predicted by semiclassical arguments \cite{Emin1976, Kabanov1993}, 
and it is in agreement with the occurrence of a level anti-crossing between localized and 
delocalized ground states in the energy spectrum \cite{Ku2007}. 

Even though the simple Holstein model is not directly applicable to many experiments, the 
coexistence of long-lived polarons and metastable delocalized states may be qualitatively
correct for systems which at the moment do not allow for a simple modeling. In fact, the 
coexistence of a Drude peak and polaronic features in photo-excited states has been 
observed in optical experiments on TaS$_2$ \cite{Dean2011}. If delocalized states are stabilized 
by an energy barrier, this suggests unique ways to control the properties of photo-excited 
states: The number of mobile carriers may be modified by second pulse that helps to bring 
electrons over the barrier, either by field-localization of the electrons, which can transiently 
increase the electron-lattice effects \cite{Werner2014}, or by exciting the delocalized carriers. 
In this way the carrier mobility could be {\em lowered} by a 
pulse, allowing for a controlled switch-on of a metallic state (by photo-exciting carriers), followed 
by a {\em switch-off} (by localizing carriers). Such possibilities will be investigated in future work. 
From a technical perspective, we note that the structure of the DMFT equations in equilibrium (a 
continued fraction) is similar to  the momentum averaged technique \cite{Berciu2006}. Hence 
the Green's function formalism presented in our work can be directly applied to extend the 
latter approach to nonequilibrium, which would be a promising way to study the time-resolved 
optical conductivity of the transient state in finite dimensions \cite{Goodvin2011}.

\acknowledgments
We thank J. Bon\v{c}a, U. Bovensiepen, D. Gole\v{z}, Z.~Lenar\v{c}i\v{c}, P. Prelov\'sek, Ph.~Werner, and L.~Vidmar  
for fruitful discussions. ME acknowledges the Aspen Center for Physics and the NSF Grant No.~1066293 for 
hospitality during writing of the manuscript. The calculations were run in part on the supercomputer HLRN
of the North-German Supercomputing Alliance.

\appendix

\section{Nonequilibrium DMFT for the Holstein model}\label{ap1}

\subsection{Nonequilibrium DMFT for the Holstein model}

\begin{figure}[tbp]
\centering
\includegraphics[width=0.9\columnwidth]{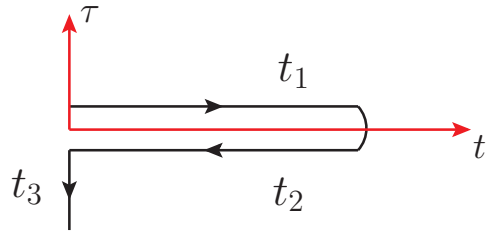}
\caption{The Keldysh contour, ranging from time $0^+$ on the upper real-time branch to 
some maximum time $t_\text{max}$, back to $0$, and to the imaginary time $-i\beta$.
Arrows indicate the direction of contour ordering.
We will use the notation $t\succ t'$ ($t\prec t'$) is $t$ is later (earlier) on $\mathcal{C}$
than $t'$. In the figure,  $-i\beta \succ t_{3} \succ t_{2} \succ t_{1} \succ 0^+$.}
\label{fig:keld}
\end{figure}

In this appendix we present the exact solution of the nonequilibrium DMFT equations for the Holstein model
in the low density limit. For an introduction to the Keldysh formalism, as well as the notation for Keldysh 
Green's functions, self-energies and Dyson equations, we refer to Ref.~\cite{Aoki2014}. Nonequilibrium 
DMFT for the Holstein model has been discussed in Ref.~\cite{Werner2013phonon}, so we will only briefly outline 
the general formalism and then focus on the low-density limit. In DMFT, the lattice model Eq.~\eqref{eq:Hol1} 
is mapped onto a single impurity model with action 
\begin{align}
\label{action}
\mathcal{S}
=
-i
\int_\mathcal{C} \!\!dt 
\big[
H_\text{loc}(t)-\mu c^\dagger c]
-i
\!\!
\int_\mathcal{C} dt dt'\,
c^\dagger(t) \Delta(t,t')c(t')
\end{align}
on the Keldysh contour $\mathcal{C}$ (see Fig.~\ref{fig:keld}).
The action describes coupling of one ``Holstein atom'' to a bath of non-interacting electrons 
by the hybridization function $\Delta(t,t')$ ($\mu$ is the chemical potential). From the  impurity 
problem one obtains the local contour-ordered Green's functions 
\begin{equation}
\label{gloc}
G(t,t') = -i \frac{1}{Z}\text{Tr} [ T_\mathcal{C} e^{\mathcal{S}}  c(t) c^\dagger(t') ].
\end{equation}
The hybridization function is determined self-consistently; we will use the closed form self-consistency 
relation $\Delta(t,t')=G(t,t')$, which corresponds to a Bethe lattice with a semi-elliptic density 
of states $D(\epsilon) =  \sqrt{4-\epsilon^2}/\sqrt{2\pi}$.
In general, the action with the hybridization function is equivalent to
an Anderson impurity Hamiltonian $H_\text{imp}=H_\text{loc} + \sum_{p} (\epsilon_p-\mu) 
a_{p}^\dagger a_p + \sum_p [V_p(t) a_{p}^\dagger c + h.c.]$ in which the impurity site is 
coupled to a number of bath orbitals $p$ with a certain choice of the parameters $V_p(t)$ 
and $\epsilon_p$, i.e., $G$ may also be computed with action $\mathcal{S}=-i\int_\mathcal{C} 
dt (H_\text{imp}(t) - \mu N)$.  (For the mapping in non-equilibrium, see Ref.~\cite{Gramsch2013}).
In the following discussion we will switch between the action and the Hamiltonian notation 
as appropriate. The final result can always be written in terms of the hybridization function,
so the precise time-dependence of the parameters  $V_p(t)$ is not needed.

\subsection{Green's functions in the low density limit}

To implement the low density limit, one takes the limit $\mu\to-\infty$, keeping only leading terms in the 
fugacity $\xi=e^{\beta\mu}$. With the equivalent impurity action, the contour-ordered Green's function
\eqref{gloc} can be  written as 
\begin{align}
G(t,\tp )=&-ie^{(it-i\tp )\mu} 
\frac{ \mathrm{Tr}[e^{\beta\mu N}\mathcal{T_{\mathcal{C}}} e^{-i\int_{\mathcal{C}}d\overline{t}
H_\text{imp}(\overline{t})}c(t)c^{\dagger}(\tp )]  }
{\mathrm{Tr}[e^{\beta\mu  N}
 \mathcal{T_{\mathcal{C}}}e^{-i\int_{\mathcal{C}}d\overline{t}H_\text{imp}(\overline{t})} ]},\label{eq:Green2}
\end{align}
by taking the term $\mu N$, which commutes with $H_\text{imp}$, out of the integral. 
One can then perform an expansion in powers of $\xi$ 
by separating the trace in contributions from $N=0,1,...$ particles,
 $\mathrm{Tr}[e^{\beta{\mu}N} \cdots] =  \sum_{n=0}^{\infty}\xi^{n}\mathrm{Tr}_{N=n}[\cdots]$.  
The result is written as
\begin{align}
 G(t,\tp )=&e^{(it-i\tp ){\mu}} \times \nonumber\\
  &[\Theta_{\mathcal{C}}(t,\tp )+\xi\Theta_{\mathcal{C}}(\tp ,t)][\widetilde{G}(t,\tp )+\mathcal{O}(\xi)],
\label{projectedAnsatz}
\end{align}
where the factor$[\Theta_{\mathcal{C}}(t,\tp )+\xi\Theta_{\mathcal{C}}(\tp ,t)]$  takes care of the fact that the 
leading contribution is $1$ if $t$ is later than $\tp$ on $\mathcal{C}$ and $\xi$ otherwise, and $\widetilde{G}(t,\tp )$ 
is given by
\begin{align}
\widetilde{G}(t,\tp )
&=
-\frac{i}{Z_{0}}\mathrm{Tr}_{N=0}[\mathcal{U}(-i\beta,t)c\,\mathcal{U}(t,\tp )c^{\dagger}\mathcal{U}(\tp ,0)],\label{eq:Green2>}
\\
\widetilde{G}(t,\tp )&=
\frac{i}{Z_{0}}\mathrm{Tr}_{N=1}[\mathcal{U}(-i\beta,\tp )c^{\dagger}\,
\mathcal{U}(\tp ,t)c\,\mathcal{U}(t,0)],\label{eq:lesonesec}
\end{align}
for $t\succ t'$ and $t\prec t'$, respectively; 
$\mathcal{U}(t,\tp )$ is the standard time-evolution operator which is given by the equation
\begin{equation}
i\partial_t \mathcal{U}(t,\tp ) = H_\text{imp}(t) \mathcal{U}(t,\tp); \,\,\, \mathcal{U}(t ,t)=1.
\end{equation}
Because the normalization factor $Z_1/Z_0$ is the average particle number $n(\mu)$
to leading order in $\xi$, one can see that the propagators \eqref{ggtrtilde} and \eqref{glestilde} are 
given by Eqs.~\eqref{eq:Green2>} and \eqref{eq:lesonesec}, respectively.

In the following we refer to the function $\widetilde G(t,t')$ which contains the leading terms of $G$ in the low-density limit as the projected Green's function.
 Before discussing the interacting case, it is useful to have a brief look at
 properties of the projected Green's function $\widetilde G_0$ in the noninteracting case ($g=0$).
 For the action \eqref{action}, $G_0$ is given by the standard Dyson equation $(i\partial_t +\mu-\epsilon_f)G_{0}(t,t') - \Delta \ast G_{0} (t,t') = \delta_\CC(t,t')$ on the contour $\mathcal{C}$,
 where $C=A\ast B$ is the convolution $C(t,t') = \int_\mathcal{C} ds A(t,s)B(s,t')$.
 By using the ansatz  \eqref{projectedAnsatz} for $G_0$ and $\Delta$ one can show that the projected functions $\widetilde G_0$ and $\widetilde \Delta$ satisfy the integral-differential equation
\begin{align}
\label{prdys}
(i\partial_{t}-\epsilon_{f})\widetilde G_{0}-[\widetilde{\Delta}\circlearrowright\widetilde G_{0}](t,\tp )=0,
\end{align}
to be solved for $t\succ t'$ with an initial condition $\widetilde G_{0}(t,t)=-i$, where 
$C=A\circlearrowright B$ is the cyclic convolution, i.e., the convolution integral $C(t,t') = \int_{\mathcal{C},t'}^t ds A(t,s)B(s,t')$
is restricted to the range in which the variables $t',s,t$ appear along $\mathcal C$ in clock-wise order,
\begin{align}
\int_{\mathcal{C},t}^{t'} ds f(s)
&=\int_{t'\succ s\succ t} 
\hspace*{-6mm}ds f(s) 
&
\hspace*{4mm} t'\succ t,
\\
\int_{\mathcal{C},t}^{t'} ds f(s)
&=
\int_{t'\succ s\succ -i\beta}
\hspace*{-10mm}ds f(s) + 
\int_{0^+\succ s\succ t}
\hspace*{-8mm}ds f(s)
&
\hspace*{4mm} t\succ t'. 
\end{align}
The Dyson equation for the projected functions $\widetilde G(t,t')$ has been derived and 
discussed in great detail in 
relation to the nonequilibrium generalization of the 
non-crossing approximation Ref.~\cite{Eckstein2010nca}; the latter can also be obtained 
as the low-density limit of a (pseudo-particle) theory, and hence the mathematical 
structure of the Green's function theory is the same. The numerical solution of the integral 
equation is also discussed in Ref.~\cite{Eckstein2010nca}.

\subsection{The interacting case}

To obtain a solution for the interacting projected Green's function $\tilde G$
we insert an identity $1=\sum_p |p \rangle\langle p|$ for the $N=0$ electron sector in Eqs.~\eqref{eq:Green2>} and \eqref{eq:lesonesec} to the 
left
of the annihilation operator $c$; in the $N=0$ sector, a full basis is given by the phonon-number states $| p \rangle =\frac{(b^{\dagger})^{p}}{\sqrt{p!}} | 0\rangle_{\mathrm{ph}}| 0 \rangle_{\mathrm{e}}$. 
To re-group the resulting terms, 
it is convenient to introduce a cyclic propagator,
\begin{align}\label{eq:u2<}
 \mathcal{U}_c(\tp ,t) 
 &=
 \left \{
 \begin{array}{ll}
 \mathcal{U}(\tp ,t) & t\prec t'
 \\
 \mathcal{U}(t',0^+) \mathcal{U}(-i\beta ,t) & t'\prec t
 \end{array}
 \right.
 \\
 &=
 \mathcal{T}_{\mathcal{C}} \exp\Big(
 -i \int_{\mathcal{C},t}^{\tp} ds \,H_\text{imp}(s)
 \Big),
\end{align}
and auxiliary propagators
\begin{align}
\mathcal{G}_{pp'}(t,\tp )
&=
-i
[
\Theta_\mathcal{C}(t,t')
-
\Theta_\mathcal{C}(t',t)
]
\langle p| c\,\mathcal{U}_c(t,\tp ) \,c^{\dagger} | p'\rangle.
\end{align}
With these definitions it is straightforward to re-group 
Eqs.~\eqref{eq:Green2>} and \eqref{eq:lesonesec}
into
\begin{align}\label{eq:Gtild_Gsmall}
 \widetilde{G}(t,\tp )=
 \frac{1}{Z_0}
 \sum\limits_{p=0}^{\infty}
 \langle p |\mathcal{U}_c(\tp, t ) | p\rangle
 \mathcal{G}_{pp}(t,\tp ).
\end{align}
The factor $\langle p | \mathcal{U}_c(\tp ,t) | p\rangle$ satisfies 
\begin{equation}
\label{upp}
 \langle p | \mathcal{U}_c( t,\tp) | p'\rangle=\delta_{pp'}
  e^{-i(t-\tp )p\omega_{0}}
  [
  \Theta_{\mathcal{C}}(t,t')
  +
  \Theta_{\mathcal{C}}(t',t) e^{-\beta p \omega_0}
  ].
\end{equation}
The next step is to derive the Dyson equation for $\mathcal{G}$. For this purpose, we evaluate 
$\mathcal{G}$ by expanding $\mathcal{U}_c$ in terms of the electron-phonon interaction 
$H_{\mathrm{ep}}=g(t)c^{\dagger}c(b+b^{\dagger})$, 

\begin{multline}
 \mathcal{G}_{pp'}(t,\tp )=
 \sum\limits_{n=0}^{\infty}(-i)^{n+1}
 \int_{\mathcal{C},\tp }^{t}\!\!dt_{1} 
 \int_{\mathcal{C},\tp }^{t_{1}}\!\!dt_{2} \cdots 
 \int_{\mathcal{C},\tp }^{t_{n-1}}\!\!dt_{n} \,\times 
  \\
   \times 
   \langle p | \mathcal{T}_{\mathcal{C}} 
   e^{-i\int_{\mathcal{C}}ds H_0(s)}
 c\,H_{\mathrm{ep}}(t_{1}) \cdots H_{\mathrm{ep}}(t_{n})c^{\dagger}  | p'\rangle.
 \label{eq:g>2}
\end{multline}
Here $H_0$ is the noninteracting part of $H_\text{imp}$, and the operator $H_{\mathrm{ep}}$
acts on the one-electron sector, in which it can be expressed in terms of phonon number states,
\begin{align}
 H_{\mathrm{ep}}
&=\sum\limits_{mm'}c^{\dagger}| m\rangle X_{mm'}(t)\langle m'| c,
\end{align}
where
$X_{mm'}=g(t)\langle m | b+b^{\dagger}| m'\rangle$ is given by
\begin{align}
 X_{mm'}
 &=g(t)\sqrt{m+1}\delta_{m',m+1}+g(t)\sqrt{m}\delta_{m',m-1}.
\end{align}
Hence we have
\begin{multline}
 \mathcal{G}_{pp'}(t,\tp )
 =
 \sum\limits_{n=0}^{\infty}(-i)^{n+1}
 \int_{\mathcal{C},\tp }^{t}dt_{1} \int_{\mathcal{C},\tp }^{t_{1}}dt_{2} \cdots \int_{\mathcal{C},\tp }^{t_{n-1}}dt_{n} 
 \\
 \sum\limits_{m_{1},m_{1}',\cdots,m_{n},m_{n}'} 
  \mathcal{G}_{0,pm_{1}}(t,t_{1})X_{m_{1}m_{1}'}\mathcal{G}_{0,m'_{1}m_{2}}(t_{1},t_{2})
  \cdots 
  \\
\times \,\,  X_{m_{n}m_{n}'}(t_{n})\mathcal{G}_{0,m'_{n}p'}(t_{n},\tp )   ,\label{eq:g>3}
\end{multline}
where $\mathcal{G}_{0}$ is the noninteracting resolvents. 
Eq.\eqref{eq:g>3} could be shortened into
\begin{align}
 \mathcal{G}&=\mathcal{G}_{0}+\mathcal{G}_{0}\circlearrowright X\circlearrowright\mathcal{G}_{0}+\mathcal{G}_{0}\circlearrowright
 X\circlearrowright\mathcal{G}_{0}\circlearrowright X\circlearrowright\mathcal{G}_{0}\cdots \nonumber \\
&=\mathcal{G}_{0}+\mathcal{G}_{0}\circlearrowright X \circlearrowright \mathcal{G}\label{eq:Dyson_g},
\end{align}
where all objects are matrices,
and $C=A\circlearrowright B$ is the cyclic convolution defined above.

This matrix-integral equation has to be solved for the diagonal elements $\mathcal{G}_{pp}$. 
Before doing this, we look at the noninteracting resolvent $\mathcal{G}_{0}$. Since electrons and phonons decouple,
 one can see from the definition that $\mathcal{G}_{0}$ is
 the product $\mathcal{G}_{0}(t,t') = \widetilde G_{0}(t,t')  \langle p | \mathcal{U}_c(t,t') | p'\rangle$,
 where $\widetilde G_{0}(t,t')$ is the projected Green's function for the pure electrons, which 
 satisfies the projected Dyson equation \eqref{prdys}. 
Because $ \langle p | \mathcal{U}_c(t,t') | p'\rangle$ is just an exponential factor [cf.~Eq.~\eqref{upp}], 
it is easy to show that 
\begin{align}
 (i\partial_{t}-\epsilon_{f}-p\omega_{0})\mathcal{G}_{0,pp'}-[\widetilde{\Delta}_{p}\circlearrowright\mathcal{G}_{0,pp'}](t,\tp )=0
\end{align}
with the initial condition $\mathcal{G}_{0,pp'}(t,t)=-i\delta_{pp'}$, where 
\begin{equation}
 \widetilde{\Delta}_{p}(t,\tp )=\widetilde{\Delta}(t,\tp )\langle p | \mathcal{U}_c(t,\tp )| p \rangle.
\end{equation}
Hence we can bring Eq.\eqref{eq:Dyson_g} to 
differential form 
by acting with $\mathcal{G}_0^{-1}$ from the left, 
\begin{align}
  (i\partial_{t}-\epsilon_{f}-p\omega_{0})\mathcal{G}_{pp'}-&[\widetilde{\Delta}_{p}\circlearrowright\mathcal{G}_{pp'}](t,\tp ) \nonumber \\
-&\sum\limits_{a=\pm1}X_{p,p+a}(t)\mathcal{G}_{p+a,p}(t,\tp )=0,\label{eq:Dyson2_g}
\end{align}
with the boundary condition $\mathcal{G}_{pp'}(t,\tp )=-i\delta_{pp'}$. This equation has a tridiagonal structure.
Like any matrix equation of that form, the equations for diagonal elements can be derived in recursive form.
(For example, a similar recursive structure is solved in the non-equilibrium variant of inhomogeneous DMFT \cite{EcksteinLayer}.)
We summarize the results:
\begin{align}
&(i\partial_{t}-\epsilon_{f}-p\omega_{0})\mathcal{G}_{pp}
\nonumber\\
&-[(\widetilde{\Delta}_{p}+\widetilde{A}_{p}+\widetilde{B}_{p})\circlearrowright\mathcal{G}_{pp}](t,\tp )=0,\label{eq:dyson_last1}
\end{align}
\begin{align}
&\widetilde{A}_{p}(t,\tp )=pg(t)\widetilde{G}_{p-1}^{[p]}(t,\tp )g(\tp ),\label{eq:dyson_last2}\\
&(i\partial_{t}-\epsilon_{f}-(p-1)\omega_{0})\widetilde{G}_{p-1}^{[p]}(t,\tp )
\nonumber\\
&-[(\widetilde{\Delta}_{p}+\widetilde{A}_{p-1})\circlearrowright\widetilde{G}_{p-1}^{[p]}](t,\tp )=0,\label{eq:dyson_last3} 
\end{align}
\begin{align}
&\widetilde{B}_{p}(t,\tp )=(p+1)g(t)\widetilde{G}_{p+1}^{[p]}(t,\tp )g(\tp ),\label{eq:dyson_last4}\\
&(i\partial_{t}-\epsilon_{f}-(p+1)\omega_{0})\widetilde{G}_{p+1}^{[p]}(t,\tp )
\nonumber\\
&-[(\widetilde{\Delta}_{p+1}+\widetilde{B}_{p+1})\circlearrowright\widetilde{G}_{p+1}^{[p]}](t,\tp )=0, \label{eq:dyson_last5}
\end{align} 
where initial conditions are $\mathcal{G}_{pp}(t,t)=\widetilde{G}_{p\pm1}^{[p]}(t,t)=-i$.
Solving Eqs.(\ref{eq:dyson_last1}, \ref{eq:dyson_last2}, \ref{eq:dyson_last3}, \ref{eq:dyson_last4}, \ref{eq:dyson_last5}), consistently 
and plugging the solution into Eq.\eqref{eq:Gtild_Gsmall} by tuning $\epsilon_f$ such that the $-i\widetilde{G}^{<}(t,t)=1$,
enable us to come up with an exact numerical solution of a single-polaron problem.

Phonon-occupation numbers [Eq.~\eqref{dist_ph}] can be read off directly from the Green's function, 
$P(p) = i\mathcal{G}_{pp}(t_+,t_-)/n_{el}Z_0$, where $n_{el}=i\widetilde G(t_+,t_-) = \sum_{p}i\mathcal{G}_{pp}(t_+,t_-)$,
cf. Eq.~\eqref{eq:Gtild_Gsmall}.
Momentum resolved Green's functions $G_{\bm k}$ as a 
function of the (time-independent) band energy $\epsilon_{\bm k}$ are 
obtained from the local Green's functions and the hybridization function by the DMFT equations as described in Ref.~\cite{Aoki2014},
but reduced to low-density, i.e., we solve the equations $\widetilde G = \widetilde Z+\widetilde Z \circlearrowright \widetilde \Delta \circlearrowright \widetilde G$ for $\widetilde Z$, and then $\widetilde G_{\bm k} = \widetilde Z+ \epsilon_{\bm k} \widetilde Z \circlearrowright \widetilde G_{\bm k}$ for $G_{\bm k}$.
 
In equilibrium, computing the propagator $\widetilde G^>$ is sufficient, because $\widetilde G^>$ and $\widetilde G^<$ are 
related by a fluctuation dissipation relation; from a Lehmann representation of Eqs.~\eqref{ggtrtilde} and \eqref{glestilde}
one can see that their spectral representations $A^{>,<}(\omega) = \pm i  \int dt  e^{i\omega t} G^{>,<}(t,0)$ satisfy
\begin{equation}
\label{FDT}
A^<(\omega) = \mathcal{N} e^{-\beta \omega}  A^>(\omega),
\end{equation}
where the normalization $\mathcal{N}$ ensures $\int A^<(\omega) d\omega =1$. Hence 
Ciuchi et al. \cite{Ciuchi1997} obtained an exact solution for $\widetilde G^>$ in terms of 
resolvents $\mathcal{G}_{pp'}^>(t-t')$, which is solved by Laplace transformation.

\section{Atomic limit}
\label{ap2}

\subsection{Lang-Firsov transformation}

Here we consider the atomic limit of the Holstein model. The Hamiltonian is given by
\begin{align}
\label{Holstein atom}
H
=
\hat n \big[g(t)(b^\dagger + b)-\mu\big]
+
\omega_0 b^\dagger  b,
\end{align}
with $\hat n=c^\dagger c $. We formally allow for a coupling $g(t)$ with arbitrary time-dependence,
of particular interest will be the case of a sudden quench, with $g(t)=g_0$ for $t\le0$ 
and $g(t)=g_1$ for $t>0$. We will compute the spectral function of the electron and the 
phonon distribution function.

To decouple electron and phonon degrees of freedom, we use the Lang-Firsov (LF) transformation,
which is a basis change that introduces a time-dependent shift of phonon coordinate. For a general 
time-dependent  LF transformation we make the ansatz
\begin{align}
\label{LFansatz}
W(t)
&=e^{i[PX_0(t)+XP_0(t)]},
\end{align}
and $\bar A= W^{\dagger}(t) A W(t)$ will denote unitary transformation of operators $A$.
Here  $X=(b^\dagger+b)/\sqrt{2}$ and $P=i(b^\dagger-b)/\sqrt{2}$ are  phonon position and 
momentum ($[X,P]=i$), and $X_0(t)$ and $P_0(t)$ depend on the electron operator only,
such that $\bar X = X-X_0$ and $\bar P=P+P_0$. This transformation was used in Ref.~\cite{Werner2013phonon}
to derive the strong-coupling solver for the Hubbard-Holstein
 model, where it was constructed 
so that electron and phonon parts decouple: When $X_0(t)$ and $P_0(t)$ satisfy the classical equations of motion 
\begin{align}
\label{eomX}
X_0'
&=
-\omega_0P_0,
\\
\label{eomP}
P_0'
&=
\omega_0X_0 -f(t),
\end{align}
with a force $f(t)=\sqrt{2} g(t)\hat n$, then the Hamiltonian for the new basis is 
\begin{align}
H_{LF} 
&=
\frac{\omega_0}{2}(X^2+P^2)
-\hat n(\mu+\lambda(t)),\label{eq:HLF}
\\
\lambda(t)
&=
\frac{g(t)g(0)}{\omega_0}
\cos(\omega_0t)
+
g(t) \int_0^t d\bar t \sin[\omega_0(t-\bar t)] g(\bar t).\label{lamquench}
\end{align}
(Note that for a time-dependent basis transformation, terms $\dot W^\dagger W$
have to included in the Hamiltonian in addition to $W^\dagger H W$).
For the transformation we get
\begin{align}
W&
=
e^{i(PX_0+XP_0)} 
=e^{[b (X_0+iP_0) -b^\dagger (X_0-iP_0)]/\sqrt{2}}.
\end{align}
We can now use that the operator $\hat n$ is time-independent in the transformation, so that
it can be taken out of the integral for $X_0+iP_0$,
\begin{align}
\label{wfinal}
W
&=
e^{[b\gamma^*(t)-b^\dagger\gamma(t)] \hat n},
\\
\gamma(t)
&=
\frac{g(0)}{\omega_0}
e^{-i\omega_0 t}
+i
\int_0^t 
d\bar t
e^{-i\omega_0(t-\bar t)}
g(\bar t).
\end{align}
For the quench \eqref{gquench},
\begin{align}
&\gamma(t) = \frac{g_1}{\omega_0}-\frac{g_1-g_0}{\omega_0}e^{-i\omega_0t}.
\label{gammaquench}
\end{align}
The transformed electron operators read, using Eq.~\eqref{wfinal},
\begin{align}
\bar c 
&=
W^\dagger(t) 
c
W(t)
=
c\,
e^{b\gamma^*(t)-b^\dagger\gamma(t)},
\\
\bar c^\dagger
&=
W^\dagger(t) 
c^\dagger
W(t)
=
c^\dagger\,
e^{b^\dagger\gamma(t)-b\gamma^*(t)}.
\end{align}

\subsection{Phonon distribution function}

The phonon distribution function 
$P(m)$ is obtained from the expectation value of the projector $|m\rangle\langle m| $, 
where $|m\rangle = \frac{1}{\sqrt{m!}} (b^\dagger)^m |0\rangle$ is the $m$-phonon state. After 
the Lang-Firsov transformation, we must evaluate the expectation 
value of
$W^\dagger(t) P_m W(t)$
in the free boson model \eqref{eq:HLF} with one electron, i.e.,
\begin{align}
P(m)
&=
\sum_{l=0}^\infty
\frac{e^{-\beta l \omega_0}}{Z_{ph}}
\big|\langle l | e^{ b \gamma^*(t)-b^\dagger\gamma(t)]} |m\rangle \big|^2.
\end{align}

Of particular interest will be the case when the system is initially not in a thermal state,
but in some given eigenstate of the displaced operator. To obtain these excited state ($l>0$)
or ground state ($l=0$) distribution function, the sum is restricted to one term,
\begin{align}
\label{ph-number}
P_l(m)
&=
\big|\langle m | e^{ b \gamma^*(t)-b^\dagger\gamma(t)]} |l\rangle \big|^2
\\
\label{p02}
&=
e^{-|\gamma(t)|^2}
\big|\langle m | e^{-b^\dagger\gamma(t)} e^{ b \gamma^*(t)} |l\rangle \big|^2,
\end{align}
where in the second line we have used the Baker Hausdorff formula.
The ground state expectation value is thus simply a Poisson distribution 
\begin{align}
P_0(m)
=e^{ -|\gamma(t)|^2 }
\frac{|\gamma(t)|^{2m}}{m!}.
\end{align}
In the case of a sudden quench, the mean $|\gamma(t)|^2$ is an oscillating function 
of time, reflecting the oscillations of the coordinate $X$.

Distributions $P_l(m)$ 
for excited states $l>0$ will only be needed when the system is stationary in 
time, $\gamma(t)=\gamma^*(t)\equiv\gamma$.
Using again the  Baker Hausdorff formula, this we can write
$P_l(n+l)=e^{-\gamma^2}|A_{l,n}(\gamma)|^2$
 with 
$A_{l,n}(\gamma)=\langle n+l| e^{b^\dagger\gamma} e^{-b\gamma} |l\rangle$.
 The latter can easily be expanded in a 
power series,
 \begin{align}
 \label{wboson}
A_{l,n}(\gamma)
&=
\sum_{r=0}^l \frac{(-\gamma)^r }{r!} 
\langle n+l| e^{b^\dagger\gamma} b^r|l\rangle
\nonumber\\
&=
\sum_{r=0}^l \frac{(-\gamma)^r \gamma^{l+r}}{r!(l+r)!} 
\langle n+l| (b^\dagger)^{n+r} b^r |l\rangle
\nonumber\\
&=
\sum_{r=0}^l \frac{(-\gamma)^r \gamma^{n+r}}{r!(n+r)!} 
\frac{\sqrt{(n+l)!}}{\sqrt{(l-r)!}} \frac{\sqrt{l!}}{\sqrt{(l-r)!}}.
\end{align}
Hence we have
 \begin{align}
 \label{wboson}
P_{l}(l+n)&=e^{-\gamma^2}\gamma^{2n}\frac{l!}{(l+n)!}
\Big[
\sum_{r=0}^l\frac{(-1)^r \gamma^{2r}}{r!}
\begin{pmatrix} l+n \\n-r\end{pmatrix}
\Big]^2,
\end{align}
which can be written as Eq.~\eqref{laguerre}
using the series representation $L_{n}^{(\alpha)}(x)=\sum_{r=0}^n(-1)^r \frac{x^r}{r!} \begin{pmatrix} n+\alpha \\ n-r \end{pmatrix}$
of the generalized Laguerre polynomials.

\subsection{Green's functions and Photoemission spectrum}

We now compute the Green's function
\begin{align}
G(t,t')
&=
-i
\frac{1}{Z}
\text{tr}
\Big[
T_\CC
e^{-i\int_\CC d\hat t H_{LF}(t)}
\bar c(t)
\bar c^\dagger(t')
\Big],\label{gf02}
\end{align}
in particular the lesser and greater components, which are then used to compute the 
photoemission spectrum and the inverse photoemission spectrum, respectively.
Because $H_{LF}$ in Eq.~\eqref{gf02} does not couple electrons and phonons, 
the Green's function can be written as a product of a purely bosonic and 
electronic Green's function,
\begin{align}
\label{green01}
G(t,t')
&=
\bar g(t,t')
w_b(t,t')
\\
\bar g(t,t')
&=
\frac{1}{Z_{el}}
\text{tr}
\Big[
T_\CC
e^{-i\int_\CC d\hat t [-\lambda(t)-\mu]\hat n}
c(t)
c^\dagger(t')
\Big]
\end{align}
\begin{align}
w_b(t,t')
=
\frac{1}{Z_{ph}}
\text{tr}
\Big[
T_\CC
&e^{-i\int_\CC d\hat t \omega_0 b^\dagger b}\times \nonumber \\
 &\times e^{b\gamma^*(t)-b^\dagger\gamma(t)}\, 
e^{b^\dagger\gamma(t')-b\gamma^*(t')}
\Big].
\end{align}

The real-time components of the electronic Green's take the usual form,
\begin{align}
\label{g0les}
\bar g^<(t,t')
&=
i f_\beta(-\mu-\lambda(0))  e^{i\int_{t'}^t d\bar t [\mu+\lambda(t)]},
\\
\label{g0gtr}
\bar g^>(t,t')
&=
-i \big[1-f_\beta(-\mu-\lambda(0))\big]  e^{i\int_{t'}^t d\bar t [\mu+\lambda(t)]},
\end{align} 
 where $f_\beta(x)=1/(1+e^{\beta x})$ is the Fermi function.
 For the bosonic factor $w_b$, we first evaluate the real-time dependence of the operators,
leading to
 \begin{align}
 \label{wboson}
 w_b(t,t')
=
\frac{1}{Z_{ph}}
\text{tr}
\Big[
e^{-\omega_0 \beta b^\dagger b}
T_\CC
\,\,
& e^{b(t)\gamma^*(t)-b^\dagger(t)\gamma(t)} \times \nonumber \\
&\,\,
\times e^{b(t')^\dagger\gamma(t')-b(t')\gamma^*(t')}
\Big],
\end{align}
with $b(t)=be^{-i\omega t}$ and $b^\dagger (t)=b^\dagger  e^{i\omega t}$. 
We can then use Eq.~(74) of Ref.~\cite{Werner2013phonon},
 \begin{align}
 \label{wbles}
 w_b^>(t,t')
&=
\exp\Big(
\frac{1}{2\sinh\big(\frac{\beta\omega_0}{2}\big)}
\Big\{
\gamma^*(t)\gamma(t')e^{\omega[\frac{\beta}{2}-i(t-t')]}
\nonumber\\
&+
\gamma(t)\gamma^*(t')e^{-\omega[\frac{\beta}{2}-i(t-t')]}
\nonumber\\
&-
[|\gamma(t)|^2+|\gamma(t')|^2]\cosh\big(\frac{\beta\omega}{2}\big)
\Big\}
\Big),
\\
\label{symmetry}
 w_b^<(t,t')
 &=
 w_b^>(t',t)
 =
 w_b^>(t,t')^*.
\end{align}
Further simplifications are possible for the quench \eqref{gquench}. Inserting 
Eq.~\eqref{gammaquench} we obtain, after some algebra,
 \begin{align}
 \label{wbfactor}
 w_{b,qu}^>(t,t')
=&
w_{b,eq}^>(t-t')\times\nonumber \\
&\times \exp\Big(
i
\frac{(g_0-g_1)g_1}{\omega_0^2}
\big[
\sin(\omega_0 t')
-
\sin(\omega_0 t)
\big]
\Big),
\end{align}
where 
$w_{q,eq}^{<,>}(t-t')$ are the bosonic factors for the equilibrium state
at the finite coupling (considering $\gamma^2$ as a time-independent constant.)
To compute the Green's function for the quench case, we furthermore must evaluate the 
exponential factors in \eqref{g0les} and \eqref{g0gtr}. Using Eq.~\eqref{lamquench}, we obtain
\begin{align}
e^{i\int_{t'}^t d\bar t [\mu+\lambda(t)]}
&=
e^{i(t-t')[\mu+\frac{g_1^2}{\omega}]}
\nonumber\\\exp\Big(
&i
\frac{(g_0-g_1)g_1}{\omega_0^2}
\big[
\sin(\omega_0 t)
-
\sin(\omega_0 t')
\big]
\Big)
\end{align}
Using the decomposition \eqref{green01}, the electronic Green's functions
\eqref{g0les} and \eqref{g0gtr}, and the explicit form \eqref{wbfactor}, we 
observe that the nonequilibrium factors containing the sin-terms in the 
electron and phonon factors cancel for the retarded Green's function,  
but not for the lesser Green's function. The Green's function can thus 
be written as
\begin{align}
G^>(t,t')
&=
G_{eq}^>(t-t')
\\
G^<(t,t')
&=
G_{eq}^<(t-t')
Q(t)Q^*(t')
\\
Q(t)
&=
\exp\Big(
2i
\sin(\omega_0 t)
\frac{(g_0-g_1)g_1}{\omega_0^2}\Big),
\end{align}
where $G_{eq}(t-t')$ is the equilibrium Green's function 
at coupling $g_1$ (taking $\lambda_1=g_1^2/\omega_0$)
\begin{align}
G_{eq}^>(t)
&=
-i[1-f_\beta(-\mu-\lambda_1)]
e^{it(\mu+\lambda_1)}
w_{b,eq}^>(t)
\\
G_{eq}^<(t)
&=
i f_\beta(-\mu-\lambda_1)
e^{it(\mu+\lambda_1)}
w_{b,eq}^>(-t).
\end{align}

The equilibrium Green's function is well known and has a spectral representation, which is stated in the main text~\cite{MahanBook}.
More generally, we can compute the time-translationally invariant Green's function for any state in which 
the coupling is not dependent on time and the oscillator is initially in a phonon number state $|l\rangle$. For this we 
must calculate
 \begin{align}
 \label{wboson}
 w_b^>(t,t')
&=
\langle l
|
e^{b(t)\gamma-b^\dagger(t)\gamma}
\,\,
e^{b(t')^\dagger\gamma-b(t')\gamma}
|
l\rangle.
\end{align}
We insert an identity $1=\sum_{m}|m \rangle \langle m|$ between the exponential operators and take the time-dependence from the 
operators to the eigenfunctions $|m \rangle$,  $|n \rangle$,
 \begin{align}
 \label{wboson}
 w_b^>(t,t')
&=
\sum_{m}
e^{i(l-m)\omega(t-t')}
\langle l
|
e^{b\gamma-b^\dagger\gamma}
|m \rangle \langle m|
e^{b^\dagger\gamma-b\gamma}
|
l\rangle.
\end{align}
The expectation value is identified with the number distribution Eq.~\eqref{ph-number}.
Thus we obtain the spectral form
 \begin{align}
 \label{wboson}
 w_b^>(t,t')&=\sum_{m=0}^\infty e^{-i\omega (m-l)(t-t')} P_l(m),
\end{align}
whose Fourier transform (together with the electronic contributions Eq.~\eqref{g0les} and \eqref{g0gtr}
and the symmetry \eqref{symmetry}) leads to the spectrum given by Eq.~\eqref{alesequiatom} in the main 
text.

% \bibliography{small_polaron03.bib}

\end{document}